# Spitzer's Solar System studies of asteroids, planets and the zodiacal cloud

**David Trilling[1], Carey Lisse[2], Dale P. Cruikshank[3], Joshua P. Emery[1], Yanga Fernández[4], Leigh N. Fletcher[5], Douglas P. Hamilton[6], Heidi B. Hammel[7], Alan Harris[8], Michael Mueller[9], Glenn S. Orton[10], Yvonne J. Pendleton[3], William T. Reach[11], Naomi Rowe-Gurney[5], Michael Skrutskie[12], Anne Verbiscer[12]**

*Version 2-July-2020 revised submission to Nature Astronomy, ed. Paul Woods*

[1] Department of Astronomy and Planetary Science, Northern Arizona University, Flagstaff, AZ 86011; david.trilling@nau.edu

[2] Johns Hopkins University Applied Physics Laboratory, Laurel, MD 20723

[3] Space Science and Astrobiology Division, NASA Ames Research Center, Moffett Field, CA, USA 94035

[4] Department of Physics & Florida Space Institute, University of Central Florida, Orlando, FL 32816

[5] School of Physics and Astronomy, University of Leicester, Leicester, LE1 7RH, UK

[6] Department of Astronomy, University of Maryland College Park, College Park, MD

[7] Association of Universities for Research in Astronomy, Washington, DC 20004

[8] DLR Institute of Planetary Research, 12489 Berlin, Germany

[8] Leiden Observatory, University of Leiden, 2333 CA Leiden, The Netherlands; SRON Netherlands Institute for Space Research, 9747 AD Groningen, The Netherlands

[10] NASA Jet Propulsion Laboratory, Pasadena CA 91109

[11] SOFIA Science Center, NASA Ames Research Center, Moffett Field, CA 94035

[12] Department of Astronomy, University of Virginia, Charlottesville, VA 22904

Proposed Running Title: "Spitzer's Solar System Science Legacy: Studies of the Relics of Solar System Formation & Evolution, Part 2"

Please address all future correspondence, reviews, proofs, etc. to:

Dr. David Trilling

Department of Astronomy and Planetary Sciences

Northern Arizona University

Flagstaff, AZ 86011

David.Trilling@nau.edu

# Abstract

In its 16 years of scientific measurements, the Spitzer Space Telescope performed a number of ground-breaking infrared measurements of Solar System objects. In this second of two papers, we describe results from Spitzer observations of asteroids, dust rings, and planets that provide new insight into the formation and evolution of our Solar System. The key Spitzer results presented here can be grouped into three broad classes: characterizing the physical properties of asteroids, notably including a large survey of Near Earth Objects; detection and characterization of several dust/debris disks in the Solar System; and comprehensive characterization of ice giant (Uranus, Neptune) atmospheres. Many of these observations provide critical foundations for future infrared space-based observations.

# I. Introduction

Two key questions in planetary science concern the formation and evolution of our Solar System, and understanding the processes that drive the functioning of the major planets. Both provide context for understanding exoplanetary systems as well as understanding the history of our own planetary system. To understand the formation of our Solar System, the many small bodies in our planetary system – asteroids, comets, and the like – provide the clearest view, as these objects act as cosmochemical and dynamical tracers of the processes that have sculpted our planetary system. Complementary studies of the highly dynamic major planets provide the opportunity to study the entire history of the Solar System, from formation to present day, by observing the Solar System today.

As with most science fields, advances in our understanding are driven by technological progress, and the advent of space-based infrared astronomy in the 1970s provided a new view of the Universe. The enormous potential of a cooled infrared telescope in space, which later became the Spitzer Space Telescope, was recognised by the Solar System science community at an early stage in its development. Although the initiative for the project came from NASA's Astrophysics Division, early design features included the ability to track objects moving against the fixed background sky. In order to promote the project, then called SIRTF (Space Infrared Telescope Facility), within the planetary science community, presentations were made at conferences and papers were published in journals and books[1,2]. In addition, a workshop held in 1999 brought more than 60 people together explore the prospects for the study of the Solar System and circumstellar disks with SIRTF. A substantial program that surveyed a wide variety of Solar System bodies was included in the first year of operations in space, and the resulting archived data continue to be studied. Numerous additional programs relevant to planetary science were proposed and executed throughout both the cold and warm phases of the operation of Spitzer, and a complete listing is available[3].

This paper represents Part 2 of our presentation of the Spitzer Solar System Legacy Science. In Part 1 (ref. 4) we presented a picture of the current paradigm for solar system formation along with a review of Spitzer's scientific results from studying outer solar system planetesimals and dwarf planets. In this paper, we briefly describe Spitzer's results for asteroids, dust/debris rings, and the ice giants Uranus and Neptune.

## II. Asteroids

**IIa. Near-Earth Objects.** Near Earth objects (NEOs) constitute the largest population of Solar System bodies observed by Spitzer. NEOs escaped from their source regions -- generally in the main asteroid belt -- via the Yarkovsky effect and planetary scattering[5] and therefore act as dynamical and composition tracers of small bodies throughout the Solar System. Some 3000 NEOs were observed by Spitzer, primarily in a series of large Warm Spitzer programs, observed after the onboard cryogen was exhausted. NEOs have surface equilibrium temperatures in the range 200--400 K, so in most cases the measured flux at 4.5 microns is dominated by thermal emission. Therefore, 4.5 micron fluxes can be used, in combination with absolute magnitudes in the optical, and the Near Earth Asteroid Thermal Model, NEATM (ref. 6), to derive the NEO's diameter and albedo. The beaming parameter, $\eta$, a critical part of the thermal model, can only be fitted if at least two thermal flux measurements are available; otherwise a value for $\eta$ has to be assumed, as is the case for almost all Spitzer-observed NEOs.

The sizes and albedos are fundamental physical properties of Solar System bodies. The size distributions of populations controlled by the conditions of formation as well as collisional evolution. Comparisons among populations can lead to genetic relationships of separate populations and their reservoirs. Albedo, or average surface reflectance, is linked to composition. Linking size distributions and albedos from NEOs to their source regions provides a link into the evolution of small body populations throughout the Solar System.

The composite result of the Spitzer NEO surveys is shown in Figure 1: diameter and albedo of 2204 unique NEOs. Overall, the uncertainty on any given solution of diameter and albedo is relatively large: ±20% and up to ±50%, respectively[7,8], but the size and albedo distribution properties of the NEO population as a whole can be derived given the large number statistics. There is a clear bias against the smallest, darkest (i.e., lowest albedo) NEOs, because the observed samples were chosen from optically detected NEOs. Future work includes debiasing these observational results using the results of the unbiased all-sky WISE/NEOWISE albedo distribution[9].

The derived albedos for a significant fraction (5% - 10%) of NEOs are very high – in many cases greater than 0.5 and in some cases higher than 0.8. These albedos may not be real, as NEOs are almost certainly rocky bodies with intrinsic albedos in the range of roughly 0.02 to 0.40. These high albedo solutions may be due largely to rotationally-induced lightcurve variability due to asynchronous infrared and optical observations[10]. Albedos and diameters are correlated for any individual asteroid solution, so a bad solution in one corresponds inversely to a poor solution in the other.

For a large number of NEOs, time-series photometry exists. These data sets have been used to derive the rotation periods, shapes, and intrinsic strengths for 38 individual objects[11] and the average shape of sub-kilometer NEOs[12]. Sub-kilometer NEOs appear to be more elongated than main belt asteroids that are larger than 1 km (ref. 12); whether this is a size effect or an effect of their dynamical environment will await forthcoming large surveys of sub-kilometer main belt asteroids.

Spitzer observed several individual NEOs of interest. One is (3552) Don Quixote, an NEO with a comet-like orbit, which was not previously known to exhibit comet-like behavior (i.e., a tail and/or coma). Observations at 4.5 microns showed the presence of a tail and coma, which was interpreted as evidence for CO or $CO_2$ emission[13]. Subsequent observations[14] have further shown activity in both optical and radio wavelengths. NEOs 2009 BD and 2011 MD were both observed as part of a campaign to characterize potential targets for NASA's Asteroid Redirect Mission. Although this mission has since been canceled, the Spitzer results preserve their scientific importance. For 2009 BD, the most likely solution is high thermal inertia and relatively low density, implying a collection of boulders and significant void space[15]. For 2011 MD, the density is also likely low, again implying significant porosity (in this case, more than 65%) (ref. 16). 2011 MD is the smallest individual object detected during the entire Spitzer mission (diameter around 6 meters).

Asteroid (101955) Bennu has recently been visited by NASA's OSIRIS-REx mission, and asteroid (162173) Ryugu was visited by JAXA's Hayabusa-2 mission. Both asteroids were observed by Spitzer prior to spacecraft arrivals. For Bennu, Spitzer photometry and spectroscopy was used to measure[17] a thermal inertia of 310 +/- 70 J m$^{-2}$ K$^{-1}$ s$^{-0.5}$, a value that is in quite good agreement with the spacecraft measurement, although the distribution of grain sizes on the surface of the asteroid has turned out to be very different than the prediction based on Spitzer data[18,19]. For Ryugu, a lower limit of 150 Jm-2K-1s-0.5 with values up to 700±200 J m$^{-2}$ K$^{-1}$ s$^{-0.5}$ was allowed by modeling, depending on the pole orientation adopted[20]; separately, a value of 150-300 J m$^{-2}$ K$^{-1}$ s$^{-0.5}$ was determined[21], in good agreement with the value measured by the spacecraft after arrival[22]. Finally, the size and albedo of 65 potential future spacecraft target asteroids were obtained, based on observations taken during the ExploreNEOs survey[23].

**IIb. Main Belt Asteroids**. Main-belt asteroid families provide insight into the collisional evolution of the belt; family members are thought to be fragments from an initial collision in which the parent asteroid was shattered or subject to a large impact. Dynamical studies enable family members to be identified by searching for associations in proper orbital elements. Spitzer data have formed the basis of a number of spectroscopic and photometric studies of members of main-belt asteroid families in order to probe their compositions and physical properties.

The thermal continua (3.5 – 22 µm) of 17 Karin cluster asteroids were measured in order to make the first direct measurements of their sizes and albedos and study the physical properties of their surfaces[24]. The targets were among the smallest main-belt asteroids observed at the time in the mid-infrared. NEATM-based diameters were derived ranging from 17 km for (832) Karin to 1.5 km for the smallest asteroid, with typical uncertainties of 10%. The mean albedo was found to be $p_v$=0.215 ± 0.015, very similar to that for 832 Karin itself, consistent with the view that the young Karin asteroids are closely related physically as well as dynamically. The mean albedo is lower than expected for young, fresh rocky "S-type" surfaces, suggesting that space weathering can darken main-belt asteroid surfaces on < 6 My timescales. The results are consistent with models of the formation of the Karin cluster via a single catastrophic collision only 5.8 ± 0.2 My ago[25].

One of the largest and oldest families in the main belt is the Themis family. In contrast to the Karin cluster, the Themis family consists of primitive low-albedo asteroids with high carbon content. Spectra from 5-14 microns were obtained for 8 Themis family asteroids in order to constrain their composition and thermal properties[26]. Using NEATM, the authors derived a mean albedo of $p_v$ = 0.07 ± 0.02, with a mean η value close to unity, implying that the thermal inertia is probably low. Five of the targets were found to display a 9-12 µm emission plateau suggestive of a fine grained silicate mantle embedded in a relatively transparent matrix or a "fairy-castle" dusty surface structure also observed in the spectra of Trojan asteroids[17].

A broad 10-μm emission feature was detected in 11 Themis-family and six Veritas-family (a much younger primitive family also in the outer main belt) spectra, which was attributed to fine-grained and/or porous silicate regolith[27]. Using the NEATM the authors found beaming parameters near unity and geometric albedos in the range 0.03-0.14 for both families. On the basis of a comparison with a laboratory study of the 12-μm spectral region in primitive meteorites, the authors infer that the asteroids in their sample are more similar to less aqueously altered meteorites.

Archival data for 62 members of the Hilda family was analyzed to find that small (<10 km) Hildas have a higher mean albedo, and a larger albedo dispersion, than large (>10 km) Hildas, suggesting that some of this difference may be due to delivery of outer Solar System material to the Hilda family early in the Solar System's history[28].

The main-belt asteroids 2867 Steins and 21 Lutetia, fly-by targets of the Rosetta Mission, were observed[29] using the Spitzer Infrared Spectrometer (IRS). Lutetia was found to have thermal emissivity similarities with carbonaceous chondrite meteorites, difficult to reconcile with its moderate albedo and M-type classification. For the E-type asteroid Steins, on the other hand, the similarities of its emissivity to enstatite meteorites could be confirmed.

Finally, archival data was used to study the albedo and size distribution of main belt asteroids using data that was obtained to map the galactic plane and a star formation region at 24 microns (ref. 30). The results included a broad albedo distribution and a size distribution that deviates from a constant power law at sizes less than 10 km.

**IIc. Jupiter's Trojan Asteroids.** Jupiter has millions of planetsimals co-rotating about the Sun with it in its L4 and L5 Lagrange points. Dubbed the Greeks (L4 population) and the Trojans (L5 population), it is debated as to whether these are objects formed nearby in the proto-planetary disk and trapped within the first ~100 Myrs of the Solar System's existence, or if they were injected from the Kuiper Belt into the Lagrange points 500 – 700 Myr later during the Jupiter-Saturn 2:1 resonance crossing of the Late Heavy Bombardment[31].

The first mid-infrared spectra of Trojans were taken using Spitzer/IRS (Figure 2). Three of the spectra (of Trojan asteroids 624 Hektor, 911 Agamemnon, and 1172 Aneas) surprisingly revealed silicate features at 9–12 um and 18–25 microns indicative of fine-grained silicates similar to those found in cometary dust spectra[17] but not in solid silicate laboratory specimens. This led to the suggestion that these objects are covered in a unique "fairy-castle" like structure of the surface[17]; others have argued for dense dust "comae" filling these objects' Hill spheres. Photometric lightcurves of the binary Trojan system Patroclus-Menoetius, one of the targets of NASA's Lucy mission, during its mutual occultation, were used to determine the thermal inertia in the system is quite low, suggesting fine grain material on the surface[33]. The density of the system was found to be around 1 g/cm3, which has significant implications both for the formation of these bodies and design for the upcoming Lucy mission.

## III. Dust and rings in the Solar System

**IIIa. Earth's Resonant Dust Ring.** Asteroid-asteroid collisions and dust released from cometary sublimation create the interplanetary (or zodiacal) dust cloud. Once released, the dust orbits decay due to Poynting-Robertson (P-R) drag until they fall into the Sun or assume escape trajectories as beta-meteoroids[34,35], but on the way in they can be trapped into metastable gravitational resonances with the planets. Theory predicts that the shape of the cloud of resonant particles depends on the sizes of the particles[36]; small particles that are dominated by radiation pressure effects pass through quickly and are not trapped. Larger particles can be trapped for much longer and are predicted to show a pronounced trailing clump where P-R drag counters a planet's gravitational attraction. In 2007 Spitzer passed through the structure of the Earth's resonant dust ring, verifying[37] the leading/trailing zodiacal light flux asymmetries known since the 1983 IRAS all-sky infrared survey (Figure 3). Comparing the dynamical predictions to Spitzer observations showed that the zodiacal light particles trapped in the Earth's resonant ring have radii of at least 10 microns.

**IIIb. The zodiacal dust cloud.** Spitzer observed the Solar System's zodiacal dust cloud -- small-grained material found in interplanetary space whose source is active asteroids and comets. Spitzer observed long strips through the ecliptic plane in order to resolve the structures of the zodiacal dust in detail and search for changes in structure over the nearly 20 year time difference between IRAS and Spitzer. Two "orphan" dust trails were identified, unassociated with any known comet and containing more surface area than all comet trails combined[38]. Those trails were linked back to "orphan" trails that had previously been discovered by IRAS, allowing a crude orbit determination showing the particles may have been produced by asteroid collisions within the last 100,000 yr. Larger-scale dust bands spreading across most longitudes were measured in new detail, allowing an association to be made with collisional debris associated with part of the Themis family and implying significant temporal variability in debris disks due to spikes in debris production after asteroid collisions[39].

**IIIc. Saturn's Phoebe Ring.** Phoebe is a large, irregularly shaped outer satellite of Saturn that is thought to be a captured KBO because of its retrograde orbit. Using MIPS 24μm and 70μm photometry, Spitzer discovered[40] the Solar System's largest planetary ring: Saturn's Phoebe ring (Fig. 4a). Most planetary rings lie within a few radii of their host body; source satellites continuously supply this dust, which is then lost in collisions or by radial transport. But Saturn's Phoebe ring is unique in both its size and orientation. The Phoebe ring extends from at least $100R_S$ to an astonishing $270R_S$ ($R_S$ denotes Saturn's radius, 60,330 km) and the ring has a vertical thickness, $40R_S$, (Fig. 4b) matching the vertical range of motion of Saturn's outer moon, Phoebe[41]. Therefore, unlike all other known rings, the Phoebe ring is centered on Saturn's orbital plane rather than its equator. (Another spacecraft, the Wide-field Infrared Survey Explorer, WISE[42], observed Saturn at 22 μm in 2010 and revealed, for the first time, the full radial extent of the Phoebe ring[41].) Phoebe ring particles smaller than centimeters in diameter slowly migrate inward[43,44], presumably on retrograde orbits like that of Phoebe.

Iapetus orbits Saturn interior to Phoebe, with a rotation period that is synchronous with its prograde orbital period of 79.3 days. Phoebe particles that drift inward are swept up by Iapetus, preferentially on its leading hemisphere in a symmetrical pattern centered on the apex of the satellite's orbital motion. This explains the observation that the visible albedo of the leading hemisphere of Iapetus is ten times less than that of the trailing hemisphere[45]. Spitzer's discovery of the Phoebe ring explained the origin of Iapetus' unusual appearance and thereby solved one of planetary science's long-standing mysteries[46]. Further cementing this link, hydrocarbons have been found in the near-infrared spectra of both Phoebe and the leading hemisphere of Iapetus, drawing a compositional connection between the two bodies[47,48].

## IV. Ice Giant Planets: Uranus & Neptune

The giant planets in our Solar System formed from planetesimals, like terrestrial planets, but formed so early and so far from the Sun that they were able to incorporate nebular gas directly from the protoplanetary disk. There are now more than 2000 giant planets known to orbit other stars, so this process of giant-planet formation is clearly ubiquitous, and thus our ice giants hold important clues to the process of planetary formation.

Jupiter and Saturn were too bright to be observed by Spitzer, but there were extensive programs to observe the ice giants Uranus and Neptune, allowing detailed characterization of their disk-averaged temperatures and molecular compositions to a greater degree of accuracy than ever before. In particular, little was known about the chemistry and circulation of Ice Giant stratospheres, even after the successful Voyager encounters of the 1980s. In the Uranian stratosphere, ethane ($C_2H_6$), methylacetylene ($CH_3C_2H$ or $C_3H_4$), and diacetylene ($C_4H_2$) were discovered in Spitzer 10-20 μm spectra[49]. Spitzer observations[50] revealed that the brightness of Uranus' mid-infrared emission varies considerably as the planet rotates. Subsequent analysis of Spitzer data[51] has revealed that this was due to stratospheric temperature varying with longitude due to some previously unexpected dynamical influence, leading to hydrocarbon brightness variability as large as 15% (Figure 5). Spatially resolving the stratospheric features responsible for this longitudinal variability (e.g., waves, vortices) will be a key goal for future observatories, including James Webb Space Telescope.

The Uranus data were combined to create a disk-averaged spectrum with high signal-to-noise ratio[50,52] (**Figure 6**). Despite the short-term variability (Figure 5), there was no compelling evidence for significant temporal variations in globally-averaged hydrocarbon abundances over the decade between Infrared Space Observatory and Spitzer observations, though we cannot preclude a possible large increase in the $C_2H_2$ abundance since the Voyager era. These results have implications with respect to the influx rate of exogenic oxygen species (e.g., from comet impacts), as well as for the production rate of stratospheric hazes on Uranus.

Vertical temperature profiles were derived from Spitzer Uranus data[50], and those remain the most detailed characterization of tropospheric and stratospheric properties to date. They are compatible with the stratospheric temperatures derived from the Voyager ultraviolet occultations measurements[54], but incompatible with hot stratospheric temperatures derived from the same data[55]. The Spitzer data also suggest an additional absorber such as $H_2S$ provides excess absorption seen above $H_2$ collision-induced absorption opacity (i.e., at 0.8–3.3 microns).

Like Uranus, Neptune's atmosphere is primarily composed of hydrogen, helium, and methane. However, Neptune spectroscopy prior to Spitzer revealed many higher-order hydrocarbons, products of reactions that are initiated by the photodissociation of $CH_4$: $CH_4$, $C_2H_6$, $C_2H_2$, and tentative evidence for $CH_3D$ and $C_2H_4$ (refs. 56, 57, 58, 59); HCN (ref. 60); $C_2H_4$ (ref. 61); and the $CH_3$ radical (ref. 62).

Hydrocarbon compounds provide crucial constraints for photochemical models of Neptune's atmosphere[63]. Such models seek to explain and predict species abundances via the balance between chemical production and destruction rates, as well as loss to the lower atmosphere by condensation and eddy diffusion[59,64]. The first observations of Neptune by Spitzer in 2004 produced the first discovery of methylacetylene ($CH_3C_2H$) and diacetylene ($C_4H_2$) in the planet's atmosphere, with derived 0.1-mbar volume mixing ratios of $(1.2 \pm 0.1) \times 10^{-10}$ and $(3 \pm 1) \times 10^{-12}$, respectively[65]. Unlike the case for Uranus, Neptune's spectrum did not show significant evidence for rotational infrared variability.

Finally, Spitzer observations of both Uranus and Neptune were also the first to reveal the existence of hydrogen dimers (two $H_2$ molecules bound by van der Waals forces) on both planets. This manifests as spectral structure surrounding the well-known hydrogen quadrupole lines that required the development of new quantum-mechanical ab initio models to explain[66].

## V.      Conclusions.

The contributions of Spitzer to Solar System science during its 16-year mission were many and varied. It is, of course, impossible to list all of Spitzer's Solar System observations and results in this short review. Spitzer's contributions to planetary science are still ongoing and expanding today, due to utilization of the science data archive[67]. Many different investigations will serve as foundational for future telescopic studies from the ground and space as well as future spacecraft explorations of our Solar System. The legacy of Spitzer's observations of bodies in the Solar System will continue even though the active mission has ended.

## VI. End notes

Correspondence and requests for materials should be addressed to David E. Trilling (david.trilling@nau.edu).

**Acknowledgements:** The authors would like to thank the Spitzer project, without which any of the science described above would have been possible. The dedication, competence, and excellence with which the staff of the Spitzer Science Center carried out their mission, and in particular observations of Solar System objects, is greatly appreciated, and has produced a scientific foundation that will last for decades.

This paper was improved by the suggestions made by two referees.

This work is based on observations made with the Spitzer Space Telescope, which was operated by the Jet Propulsion Laboratory, California Institute of Technology under a contract with NASA. Support for this work was provided by NASA, in some cases through an award issued by JPL/Caltech. Y. Fernández acknowledges support of a SIRTF/Spitzer Fellowship.

**Author contributions:** DET, CL, DPC, YF, LNF, DPH, HBH, AH, MM, GSO, YJP, WR, MS, NRG, and AV carried out scientific analysis and wrote parts of this paper. JPE contributed scientific analysis that is presented in this paper.

**Competing interests:** The authors declare no competing interests.

## VII. References.

1. Cruikshank, D.P., Werner, M.W., Backman, D.E. 1992. SIRTF: Capabilities for planetary science. *Adv. Space Res.* **12**, 187

2       Cruikshank, D.P., Werner, M.W. 1997. The study of planetary systems with the Space Infrared Telescope Facility (SIRTF). In *Planets Beyond the Solar System and the Next Generation of Space Missions*, David Soderblom, ed., Astron. Soc. Pacific Conference Series, **119**, 223

3       Spitzer approved programs: http://ssc.spitzer.caltech.edu/warmmission/scheduling/approvedprograms/ (2020)

4       Lisse, C.M. et al. 2020. Final title to go here. Nature Astronomy, **final volume**, final page number [this is the other paper in this series – probably published side-by-side with this paper; fill in reference when ready]

5       Granvik, M. et al. 2018. Debiased orbit and absolute-magnitude distributions for near-Earth objects. *Icarus*, **312**, 181

6       Harris, A.W. 1998. A thermal model for Near-Earth Asteroids. *Icarus*, **131**, 291

7       Harris, A.W. et al. 2011. ExploreNEOs II: The accuracy of the Warm Spitzer Near-Earth Object survey. *Astron. J.*, **141**, 75

8       Trilling, D.E. et al. 2016. NEOSurvey 1: Results from the Warm Spitzer exploration science survey of Near-Earth Object properties. *Astron. J.*, **152**, 172

9       Mommert, M. et al. 2015. ExploreNEOs VIII: Dormant short-period comets in the Near-Earth Asteroid population. *Astron. J.*, **150**, 106

10      Gustafsson, A. et al. 2019. Spitzer albedos of Near-Earth Objects. *Astron. J.*, **158**, 67

11      Hora, J.L. et al. 2018. Infrared light curves of Near-Earth Objects. *Astroph. J. Supp.*, **238**, 22

12      McNeill, A., Hora, J.L., Gustafsson, A., Trilling, D.E., Mommert, M. 2019. Constraining the shape distribution of Near-Earth Objects from partial light curves. *Astron. J.*, **157**, 164

13      Mommert, M. et al. 2014. The discovery of cometary activity in Near-Earth Asteroid (3552) Don Quixote. *Astroph. J.*, **781**, 25

14      Mommert, M. et al. 2020. Recurrent cometary activity in Near-Earth Object (3552) Don Quixote. *Planet. Sci. J.*, **1**, 12

15      Mommert, M. et al. 2014. Constraining the physical properties of Near-Earth Object 2009 BD. *Astroph. J. Lett.*, **786**, 148

16      Mommert, M. et al. 2014. Physical properties of Near-Earth Asteroid 2011 MD. *Astroph. J. Lett.*, **789**, 22

17	Emery, J.P., Cruikshank, D.P., van Cleve, J. 2006. Thermal emission spectroscopy (5.2-38 µm) of three Trojan asteroids with the Spitzer Space Telescope: Detection of fine-grained silicates. *Icarus*, **182**, 496

18	DellaGiustina, D.N. et al. 2019. Properties of rubble-pile asteroid (101955) Bennu from OSIRIS-REx imaging and thermal analysis. *Nature Astronomy*, **3**, 341

19	Rozitis, B. et al. 2020, Science Advances, in press

20	Campins, H. et al. 2009. Spitzer observations of spacecraft target 162173 (1999 JU3). *Astron. Astroph.*, **503**, L17

21	Müller, T.G. et al. 2017. Hayabusa-2 mission target asteroid 162173 Ryugu (1999 JU3): Searching for the object's spin-axis orientation. *Astron. Astroph.*, **599**, A103

22	Sugita, S. et al. 2019. The geomorphology, color, and thermal properties of Ryugu: Implications for parent-body processes. *Science*, **364**, eaaw0422

23	Mueller, M. et al. 2011. ExploreNEOs III: Physical characterization of 65 potential spacecraft target asteroids. *Astron. J.*, **141**, 109

24	Harris, A.W., Mueller, M., Lisse, C.M., Cheng, A.F. 2009. A survey of Karin cluster asteroids with the Spitzer Space Telescope. *Icarus*, **199**, 86-96.

25	Nesvorny, D. et al. 2006. Karin cluster formation by asteroid impact. *Icarus*, **183**, 296

26	Licandro, J. et al. 2012. 5-14 µm Spitzer spectra of Themis family asteroids. *Astron. Astroph.*, **537**, A73.

27	Landsman, Z.A. et al. 2016. The Veritas and Themis asteroid families: 5-14 µm spectra with the Spitzer Space Telescope. *Icarus*, **269**, 62

28	Ryan, E.L, Woodward, C.E. 2011. Albedos of small Hilda group asteroids as revealed by Spitzer. *Astron. J.*, **141**, 186

29	Barucci, M.A. et al. 2008. Asteroids 2867 Steins and 21 Lutetia: Surface composition from far infrared observations with the Spitzer Space Telescope. *Astron. Astroph.*, **477**, 665

30	Ryan, E.L. et al. 2015. The kilometer-sized Main Belt asteroid population revealed by Spitzer. *Astron. Astroph.*, **578**, A42

31	Morbidelli, A., Levison, H.F., Tsiganis, K., Gomes, R. 2005, Chaotic capture of Jupiter's Trojan asteroids in the early Solar System. *Nature*, **435**, 462

32	Kelley, M.S.P., Woodward, C.E., Gehrz, R.D., Reach, W.T., Harker, D.E. 2017. Mid-infrared spectra of comet nuclei. *Icarus*, **284**, 344 33      Mueller, M. et al. 2010. Eclipsing binary Trojan asteroid Patroclus: Thermal inertia from Spitzer observations. *Icarus*, **205**, 505

34      Burns, J.A., Lamy, P.L, Soter, S. 1979. Radiation forces on small particles in the solar system. *Icarus*, **40**, 1

35      Szalay, J.R et al. 2020. The near-Sun dust environment: Initial observations from the Parker Solar Probe. *Astroph. J. Supp.*, **246**, 27

36      Dermott, S.F., Jayaraman, S., Xu, Y.L., Gustafson, B.A.S, Liou, J.C. 1994. A circumsolar ring of asteroidal dust in resonant lock with the Earth. *Nature*, **369**, 719

37      Reach, W.T. Structure of the Earth's circumsolar dust ring. 2010, *Icarus*, **208**, 848

38      Nesvorny, D. et al. 2006. Candidates for asteroid dust trails. *Astron. J.*, **132**, 582

39      Nesvorny, D. et al. 2008. Origin of the near-ecliptic circumsolar dust band. *Astroph. J. Lett.*, **679**, L143

40      Verbiscer, A., Skrutskie, M.F., Hamilton, D.P. 2009. Saturn's largest ring. *Nature*, **461**, 1098

41      Hamilton, D.P., Skrutskie, M.F., Verbiscer, A.J., Masci, F.J. 2015. Small particles dominate Saturn's Phoebe ring to surprisingly large distances. *Nature*, **522**, 185

42      Wright, E.L. et al. 2010. The Wide-field Infrared Survey Explorer (WISE): Mission description and initial on-orbit performance. *Astroph. J.*, **140**, 1868

43      Soter, S. 1974, IAU Colloq. 28.

44      Burns, J.A., Hamilton, D.P., Mignard, F., Soter, S. 1996. The contamination of Iapetus by Phoebe dust. *Astron. Soc. Pacif. Conf.*, **1048**, 179

45      Squyres, S.W., Sagan, C. 1983. Albedo asymmetry of Iapetus. *Nature*, **303**, 782

46      Tamayo, D., Burns, J.A., Hamilton, D.P., Hedman, M.M. 2011. Finding the trigger to Iapetus' odd global albedo pattern: Dynamics of dust from Saturn's irregular satellites. *Icarus*, **215**, 260

47      Clark, R.N. et al. 2012. The surface composition of Iapetus: Mapping results from Cassini VIMS. *Icarus*, **218**, 831

48      Dalle Ore, C.M., Cruikshank, D.P., Clark, R.N. 2012. Infrared spectroscopic characterization of the low-albedo materials on Iapetus. *Icarus*, **221**, 735

49      Burgdorf, M., Orton, G., van Cleve, J., Meadows, V., Houck, J. 2006. Detection of new hydrocarbons in Uranus' atmosphere by infrared spectroscopy. *Icarus*, **184**, 634 50	Orton, G.S. et al. 2014. Mid-infrared spectroscopy of Uranus from the Spitzer Infrared Spectrometer: 2. Determination of the mean composition of the upper troposphere and stratosphere. *Icarus*, **243**, 471

51	Rowe-Gurney, N. et al. 2020, in prep.

52	Orton, G.S. et al. 2014. Mid-infrared spectroscopy of Uranus from the Spitzer Infrared Spectrometer: 1. Determination of the mean temperature structure of the upper troposphere and stratosphere. *Icarus*, **243**, 494

53	Houck, J.R. et al. 2004. The Infrared Spectrograph (IRS) on the Spitzer Space Telescope. *Astroph. J. Supp.,* **154**, 18

54	Herbert, F. et al. 1987. The upper atmosphere of Uranus: EUV occultations observed by Voyager 2. *J. Geophys. Res.*, **92**, 15093

55	Stevens, M.H., Strobel, D.F., Herbert, F.H. 1993. An analysis of the Voyager 2 Ultraviolet Spectrometer occultation data at Uranus: Inferring heat sources and model atmospheres. *Icarus*, **101**, 45

56	Orton, G.S. et al. 1987. The spectra of Uranus and Neptune at 8-14 and 17-23 μm. *Icarus*, **70**, 1

57	Conrath, B.J. et al. 1989. Infrared observations of the Neptunian system. *Science*, **246**, 1454

58	Bezard, B., Romani, P.N., Conrath, B.J., Maguire W.C. 1991. Hydrocarbons in Neptune's stratosphere from Voyager infrared observations. *J. Geophys. Res.*, **96**, 18961

59	Bishop, J. et al. 1995. The middle and upper atmosphere of Neptune. In *Neptune and Triton*, D.P. Cruikshank editor, UA Press, 427

60	Marten, A. et al. 1993. First observations of CO and HCN on Neptune and Uranus at millimeter wavelengths and their implications for atmospheric chemistry. *Astroph. J.*, **406**, 285

61	Schulz, B. et al. 1999. Detection of C2H4 in Neptune from ISO/PHT-S observations. *Astron. Astroph.*, **350**, L13

62	Bezard, B., Romani, P.N., Feuchtgruber, H., Encrenaz, T. 1999. Detection of the methyl radical on Neptune. *Astroph. J.*, **515**, 868

63	Moses, J.I., Fletcher, L.N., Greathouse, T.K., Orton, G.S., Hue, V. 2018. Seasonal stratospheric photochemistry on Uranus and Neptune. *Icarus*, **307**, 124

64	Moses, J.I. et al. 2005. Photochemistry and diffusion in Jupiter's stratosphere: Constraints from ISO observations and comparisons with other giant planets. *J. Geophys. Res.*, **110**, E08001 65	Meadows, V.S. et al. 2008. First Spitzer observations of Neptune: Detection of new hydrocarbons. *Icarus*, **197**, 585

66	Fletcher, L.N., Gustafsson, M., Orton, G.S. 2018. Hydrogen dimers in giant-planet infrared spectra. *Astroph. J. Supp.*, **235**, 24

67	Spitzer Heritage Archive: https://irsa.ipac.caltech.edu/data/SPITZER/docs/spitzerdataarchives/ (2020)

# VIII. Figures captions

**Figure 1:** Composite albedo and diameter results for the Spitzer NEO surveys (2204 unique objects). Overall, the uncertainty on any given solution of diameter and albedo is relatively large (±20% and up to ±50%, respectively; refs. 7 and 8), but the size and albedo distribution properties of the NEO population as a whole can be derived given the large number statistics. There is a clear bias against the smallest, darkest (i.e., lowest albedo) NEOs, because the observed samples were chosen from optically detected NEOs.

**Figure 2:** Spitzer/IRS spectra of three Jovian D-type Trojans: (911) Agamemnon, (624) Hektor, and (1172) Aneas (from ref. 17), compared to the nucleus of comet 10P/Tempel 2 (ref. 32). The spectra are very similar and all show 9-12, 18-20, and 34 micron features indicative of fine-grained ferromagnesian olivine and pyroxene silicates, similar to features seen in cometary coma dust spectra.

**Figure 3:** Model of the Earth's dust ring based on Spitzer's measurements around the Earth's orbit. Spitzer's observations showed that a ring of dust from comet tails and broken asteroids follows the Earth in its orbit. A wide cloud of these >20 micron particles, about 7 million miles wide, trails behind the Earth at about 80 times the Earth-Moon distance. Yellow/blue/black -- low to high resonant ring dust density. Spitzer's orbital path relative to Earth is shown as the red and purple curve, with the cryogenic and warm mission periods labeled. Adapted from ref. 36.

**Figure 4 – Saturn's Phoebe ring.** (*left*) Model full-ring image based on the results of Spitzer's MIPS photometric imaging. (*middle*) Detail of the 24 um Spitzer ring imaging, showing the location of 3 transepts used to extract profile information. The ring is notably diffuse in its interior but has relatively well defined, sharp boundaries. (right) Profiles cuts through the ring, showing its relative interior uniformity, minima at and symmetry around Saturn's orbital plane, and edge boundaries. Figure adapted from ref. 39.

**Figure 5:** Longitudinal variability of Uranus in December, 2007, showing percentage radiance difference from the global average spectrum, along with non-negligible standard errors. *Top:* Significant variations exist for hydrocarbon species present in the stratosphere; the hydrogen-helium continuum and deuterated methane (arising from higher pressures in the troposphere) show considerably less variation. A 10% amplitude sinusoid is shown for reference, but has not been fitted to the sparse data. *Bottom:* Hydrogen quadrupole measurements, with a 5% amplitude sinusoid displayed for reference (but not fitted). Adapted from Rowe-Gurney et al. (2020).

**Figure 6:** Disk integrated spectrum of Uranus. *Panel A:* Radiance as a function of wavelength. Different instrument modules are coded in different colors and shapes: SL2 (orange), SL "bonus order" (purple), SL1 (green), LL2 (blue), SH (red), LH (purple). (Details on the instrument modules can be found in ref. 52.) *Panel B:* Equivalent brightness temperature as a function of wavelength (bottom axis) and wavelength (upper axis). Some spectral features are identified for clarity in each panel. (Figure adapted from ref. 53.)

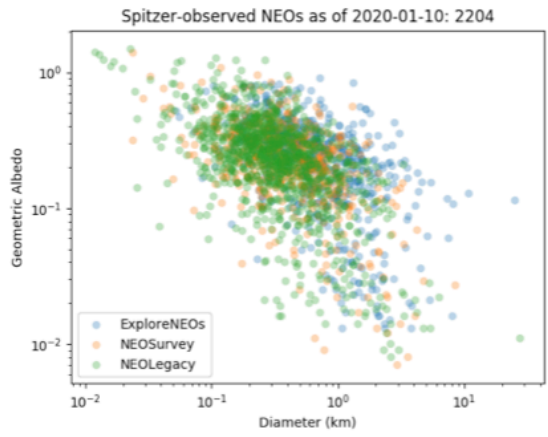

Figure 1

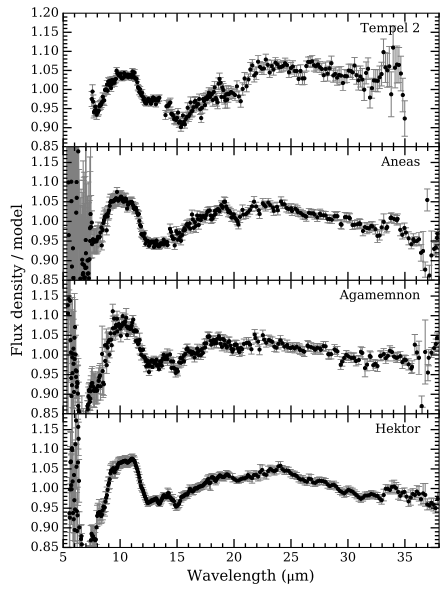

Figure 2

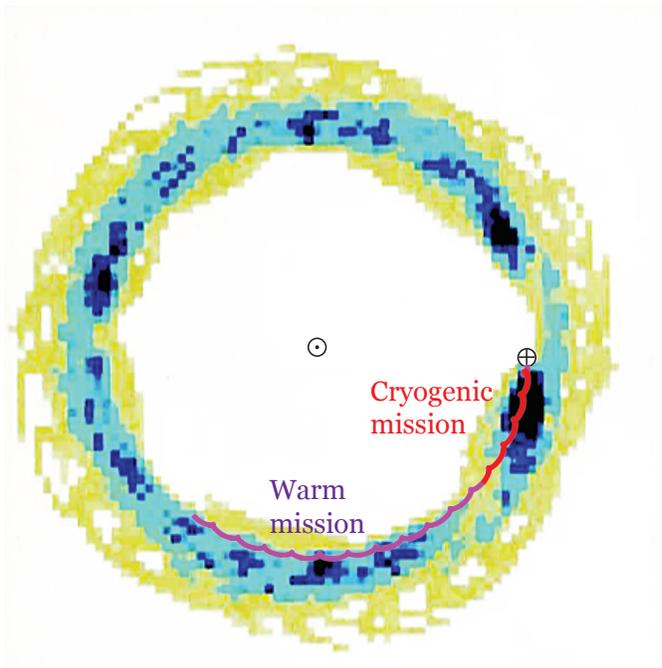

Figure 3

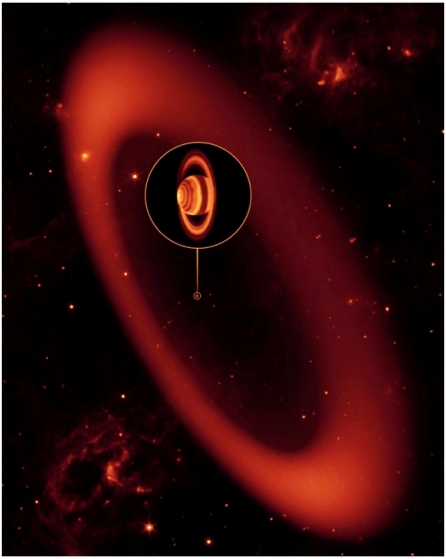

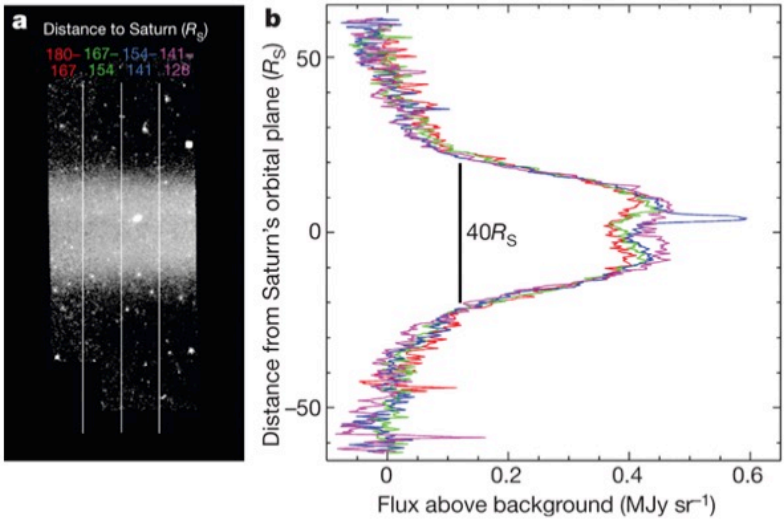

Figure 4

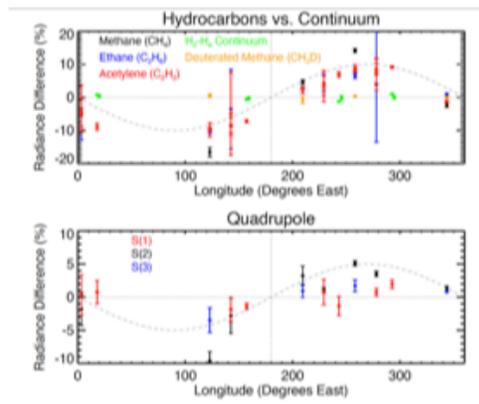

Figure 5

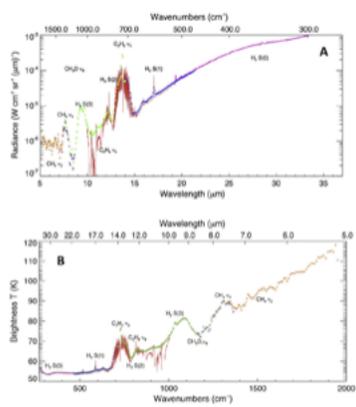

Figure 6